
         THIS IS A TEX FILE

\line{\hfill TAUP-2005-92}
\vskip .1in
\line{\hfill Oct,1992}
\date{}
\vskip 1 true cm
\titlepage
\title{\bf COSET MODELS OBTAINED BY TWISTING WZW MODELS
AND STRINGY CHARGED BLACK HOLES IN FOUR DIMENSIONS}
\vskip 1 true cm
\centerline{\caps ~DAVID ~GERSHON}
\centerline{\it School of Physics and Astronomy}
\centerline{\it Beverly and Raymond Sackler Faculty of Exact Sciences}
\centerline{\it Tel-Aviv University, Tel-Aviv 69978, Israel}
\footnote
{\dagger}{e-mail: GERSHON@TAUNIVM.BITNET}
\vskip 2 true cm
\abstract

We show that several WZW coset models can be obtained by applying
$O(d,d)$
symmetry transformations (referred to as twisting) on WZW models.
This leads to a conjecture that WZW  models gauged by $U(1)^n$
subgroup can be obtained by twisting (ungauged) WZW models.
In addition, a class of solutions that describe
 charged black holes in four dimensions is
derived by twisting $SL(2,R)\times SU(2)$ WZW action.
\endpage
\chapter{Introduction}

Symmetries play an important role in conformal field theories and
in string theories in particular.  Duality
is, for example one of the important features that reveal
some of the intrinsic properties of theories. Hence, investigation
of symmetries is very helpful on the way to fully understand string
theories.

Two major approaches are used in studying backgrounds in string
theories. The first one is by using  known exact conformal theories,
such as WZW coset models. The second approach is to study
 classical solutions, namely,
theories
with
 beta-function that vanishes to one loop order (or to a certain
order).
Yet,
when  studying unfamiliar classical solutions,
we have no way to know whether
they  correspond to exact conformal field theories.

 Recently, an interesting
  link between exact conformal theories and some classical
solutions was shown by Sen \Ref\sen {A.Sen \journal Phys.Lett.
&B271 (91) 295} \Ref \hassan {S.F. Hassan and A.Sen \journal
Nucl.Phys. &B375 (92) 103}, in terms of symmetry transformations.
  It was shown that when a
background is independent of $d$ of the $D$ target space coordinates,
the $O(d,d)$ symmetry that it has is exact to
any order in $\alpha '$- the inverse of the string tension.
 The $O(d,d)$ symmetry transformations at the classical level
were originally derived
 \Ref\venez{G. Veneziano \journal Phys.Lett. &B265 (91) 287}
\Ref \meiz {K. Meissner and G. Veneziano \journal Phys.Lett.
& B267 (91) 33} \Ref\meii{K. Meissner and G. Venezianno \journal
Mod.Phys.Lett & A6 (91) 3397}
\Ref\duf{M. Duff \journal Nucl.Phys.& B335 (90) 610}
from the effective action.
 Thus, applying
an $O(d,d)$ symmetry transformations on  backgrounds that
account for  exact
conformal theories, one is guaranteed that the new solutions,
that are referred to as the {\it twisted} solutions,
account for exact conformal theories.

The usefulness of this symmetry is large. On one hand,
in some cases it can be used to show that  a classical solution
corresponds to an exact conformal theory. On the other hand,  we
can apply
the symmetry transformations on  backgrounds which correspond  to
conformal field theories and
 find new  solutions which   account
for exact conformal theories. Such solutions were derived in few
papers
\Ref \senn {A. Sen \journal Phys.Lett. & B274 (92) 33}
\Ref \senm {A. Sen, "Macroscopic Charged Heterotic Strings",
preprint TIFR-TH-92-29}
\Ref \pat {S. P. Khastagir and J. Maharana, "Symmetries of
String Effective Action and Space Time Geometry", preprint
IP-BBSR-92-38}, and in some others.

In this paper we show another important feature of the
$O(d,d)$ symmetry. We demonstrate that several WZW coset
models can be obtained by applying  symmetry transformations
on ungauged WZW actions. Our examples lead to a conjecture that
 all WZW actions gauged by some $U(1)$ subgroups can be obtained
by symmetry transformations from (ungauged) WZW actions.

 Then we  apply the symmetry transformation on $SL(2,R)\times
SU(2)$ WZW to obtain a new class of solutions which describe
charged  black holes in
four dimension.

The paper is organized in the following way: In section 2 we
review the $O(d,d)$ symmetries and the explicit symmetry
 transformations  to one loop order.
 This contains an introductory summary only.

In section 3 we show that several coset models of WZW
gauged by $U(1)$ or $U(1)^2$ subgroups  can be
obtained directly by twisting (ungauged) WZW models.
The coset models we discuss are obtained from $G_i$ or
$G_i\times G_j$ WZW actions,
where $G_i$ is  the group $SL(2,R)$ or $SU(2)$.

In section 4 we apply  symmetry transformation on
$SL(2,R)\times SU(2)$ and
find  a new class of solutions  that describe black holes
 with electromagnetic and axionic fields in four dimensions.
These black holes carry electric charge but no magnetic
charge. Duality transformations of the electromagnetic
tensor provides solutions with magnetic charge.

\chapter{ Review of The O(d,d) Symmetry and Twisted Solutions}

Twisted solutions [\sen ,\senn]
are conformal backgrounds (described by target space metric,
 anti-symmetric tensor and dilaton field) that are obtained by
applying symmetry transformations on  conformal backgrounds.
 $O(d,d)$ symmetry
 appears when the background is independent
of $d$  of the $D$ target space coordinates,
 and   was proven
by Sen [\sen ,\hassan ],
 by means of string field theory,
to be an exact symmetry of the action, $\ie$ to hold to all orders in
$\alpha\prime$- the inverse of the string tension.
Therefore, when transforming backgrounds that correspond to
  exact conformal
field theories,
  the twisted solutions correspond  to exact
conformal field theories as well. The $O(d,d)$ symmetry transformations
include not only
duality transformations but also transformations from one conformal
field theory to another.

To one loop order, the $O(d,d)$ symmetry transformations
can be derived from the effective action.
Here
we concentrate on string theories without gauge fields
and we shall use the notations in [\senn ].
Generalization of the symmetry in the case of
the heterotic strings, is shown in [\hassan].
 Also, we shall not bring here
 the string field theory arguments, as they are not used in our
derivations.  They can be found  in $\lbrack \sen , \hassan
\rbrack$.

Let us consider a conformal background with $D$ target space coordinates
$X_{\mu}$ with $\mu =1,...,D$.
This corresponds to a space-time metric
           $g_{\mu\nu}(X)$, an anti-symmetric
tensor $b_{\mu\nu}(X)$ and a dilaton field $\Phi(X)$.
The conditions for vanishing of the $\beta$-function to one loop
order yield the one loop effective action:
$$S=\int d^Dx \sqrt{\det g} e^{\Phi}(g_{\mu\nu}\partial^{\mu}\Phi
\partial^{\nu}\Phi + R^{(D)}(g)-{1\over 12}H_{\mu\nu\rho} H^{\mu
\nu\rho}-\Lambda)\eqn\m$$
where $H_{\mu\nu\rho}=\partial _{\mu} b_{\nu\rho}+$ cyclic
permutations, $R^{(D)}$ denotes the $D$ dimensional Ricci scalar
and $\Lambda$ is the cosmological constant, equal to
$(D-26)/3$ for bosonic strings or $(D-10)/2$ for fermionic
strings.

We consider the case where both the target space metric and
the anti-symmetric tensor are independent of $d$ of the $D$ coordinates.
We refer to these coordinates as  $X_i$ where
 $i=1,...,d$, and the rest
of the coordinates as $Y_{\alpha}$ where $\alpha=d+1,...,D$.
               Now we concentrate on the
case where $g_{i\alpha}=b_{i\alpha}=0$,
$\ie$ the background has the form
$$\left (\matrix{g_{ij} & 0\cr 0& g_{\alpha\beta}\cr}\right)
\;\;\;\;\;\;\;\;
\left (\matrix{b_{ij} & 0\cr 0& b_{\alpha\beta}\cr}\right) \eqn\ml$$
Let us denote $g_{\alpha\beta}$ by $\tilde
  G_{\alpha\beta}$, which is a $(D-d)
\times (D-d)$ metric, and $g_{ij}$ by $G_{ij}$, which is a
$d\times d$ matrix. Similarly, we denote $b_{ij}$
by $B_{ij}$ and $b_{\alpha\beta}$ by $\tilde B_{\alpha\beta}$.
Then the action $\m$  takes the form
$$S=\int d^dx\int d^{(D-d)}y \sqrt{\det \tilde G} e^{-\chi} (
\tilde G_{\alpha\beta}\partial^{\alpha}
\chi\partial^{\beta}\chi+ \tilde R ^{(D-d)}(\tilde G)-
{1\over 12}\tilde H_{\alpha\beta\gamma} \tilde
H^{\alpha \beta\gamma}-\Lambda$$$$
+{1\over 8}\tilde G_{\alpha\beta} \Tr (\partial^{\alpha} ML
\partial^{\beta}M L))\eqn\mm$$ where $L$ is the $2d\times 2d$
 matrix $$L=\left(\matrix{0&{\bf 1}\cr
{\bf 1}&0\cr}\right )\eqn\matr$$
$$\chi=\Phi-{1\over 2}\ln\det  G\eqn\chi$$
and $$M=\left (\matrix {G^{-1}&-G^{-1}B\cr
BG^{-1}&G-BG^{-1}B\cr}\right)\eqn\matm$$
Let us first assume that none of the $d$ coordinates is time-like.
In this case the action is invariant under the transformation
$$M\to \Omega M\Omega^T \eqn\ome$$ where $\Omega$ is a
$2d\times 2d$ matrix, satisfying the condition $\Omega L\Omega^T=L$.
This transformation is the $O(d,d)$ symmetry of the background.
$\Omega$ corresponds to three types of transformations:
(a) Global coordinate transformations $\ie$ $X_i\to A_i^jX_j$, with
$A$ a constant $d\times d$ matrix. This corresponds to
$$\Omega=\left(\matrix {(A^T)^{-1}&0\cr 0&A\cr}\right)\eqn\amat$$
(b) Transformations of the anti-symmetric tensor in the following
form $B_{ij}\to B_{ij}+C_{ij}$, where $C$ is a $d\times d$
anti-symmetric constant matrix. This accounts for$$\Omega=\left(\matrix{
1&0\cr C&1\cr}\right)\eqn\asm$$
(c) $O(d)\times O(d)$ transformations. For any two $O(d)$ matrices
$R,S$ we can take $$\Omega={1\over 2}\left(\matrix{R+S&R-S\cr
 R-S& R+S\cr}\right )\eqn\omega$$
 When  one of the $d$ coordinates is time-like this symmetry
 is modified to $O(d-1,1)\times O(d-1,1)$. Then $\Omega$
takes the form $$\Omega= {1\over 2}\left (\matrix{
\eta (R+S)\eta& \eta (R-S)\cr (R-S)\eta& R+S\cr} \right) \eqn\ett$$
where $\eta$ is the diagonal  $d\times d$
matrix $\eta=$diag$(-1,1,...,1)$ and
$S,R$ are $O(d-1,1)$ matrices, satisfying  $S\eta S^T=R\eta R^T=\eta$.

Let us denote the matrices $A=R+S$, $C=R-S$. The symmetry transformations
$G\to G\prime$, $B\to B\prime$ $\Phi\to\Phi\prime$
for the $O(d-1,1)\times O(d-1,1)$ symmetry
are the following:
$$G\prime ^{-1}={1\over 4}\eta( A\eta G^{-1}\eta A^T+ C(G-BG^{-1}B
)C^T- A\eta G^{-1}BC^T+CBG^{-1}\eta A)\eta\eqn\metric$$
$$B\prime ={1\over 4}( C\eta G^{-1}\eta A^T+ A(G-BG^{-1}B
)C^T- C\eta G^{-1}BC^T+ABG^{-1}\eta A^T)\eta G\prime \eqn\antsy$$
$$\Phi\prime=\Phi-{1\over 2}\ln\det G+{1\over 2}\ln\det G \prime\eqn
\dilat$$
and for the $O(d)\times O(d) $ symmetry
we replace $\eta$ by the unit matrix and $R,S$  by $O(d)$
matrices.

Armed with  the $O(d,d)$ transformations, we turn to show
in the next sections some interesting applications.

\chapter{ Coset Models Obtained by Twisting WZW models}

After the $O(d,d)$
symmetry transformations were explained, we turn to show
that some  known WZW coset models can be obtained by $O(d,d)$
symmetry
transformations from (ungauged) WZW models. In other words,
the coset models
can be obtained as twisted solutions of (ungauged) WZW models.
The procedure we use is the following.  As the symmetry
transformations are
 reversible, we start with known coset models
and show
that by twisting their backgrounds
 we can obtain backgrounds that describe
(ungauged) WZW models.  We have, however to clarify two points:
Although it might look like the symmetry is applicable to
 solutions with total central charge $c=26$ (or
$c=15$ in superstrings) only, this is not the case.
 We can think of our coset models as part
of the background whereas
 the other part, which is decoupled, being another
conformal theory with the appropriate central charge (so that
the total central charge is c=26 or c=15).
 We shall not touch
this extra background, so that the symmetry can really be considered as
twisting only the coset models (with any level).
  The second point is rather obvious: Since the symmetry does not change
the number of target space coordinates, thus by twisting ungauged models
we obtain the coset models together
with the appropriate number of  decoupled
 free fields.

 In our first example
we show that gauged $SL(2,R)/U(1) $ WZW model can be obtained
by twisting $SL(2,R)$ WZW model.

The sigma-model that is described by $SL(2,R)$ WZW model
with level $k$
 can be written as \Ref\digj{R. Dijkgraaf, E. Verlinde and
H. Verlinde, preprint IASSNS-HEP-91-14}:
$$S_{SL(2,R)}={k\over 8\pi}\int d^2\sigma (4\partial_+r\partial_-r
+\partial_+\phi_L\partial_-\phi_L +\partial_+\phi_R\partial_-\phi_R
+2\cosh 2r \partial_-\phi_L\partial_+\phi_R)\eqn\a$$
where the elements of the group $SL(2,R)$ are parameterized by
$$g=e^{i{1\over 2}\phi_L\sigma_2}e^{r\sigma_1}
e^{i{1\over 2}\phi_R\sigma_2}\eqn\qa$$
We  define two fields by $$x=(\phi_L+\phi_R)/2
\;\;\;\;\;\;\;\;\;\;\;\;
y=(\phi_L-\phi_R)/2
\eqn\aqw$$ In terms of the fields $r,x,y$ the action can be written
as $$S_{SL(2,R)}=
{k\over 2\pi}\int d^2\sigma (\partial_+r\partial_-r
+\cosh^2 r \partial_+x\partial_-x
-\sinh ^2r\partial_+y\partial_-y  $$$$
+{1\over 2}\cosh 2r(\partial_+x\partial_-y- \partial_-x\partial_+y))
\eqn\aqq$$
This describes the sigma-model metric
$$G_{rr}=1,\;\;\; G_{xx}=\cosh^2 r,\;\;\;G_{yy}=-\sinh^2r\eqn\ap$$
and the anti-symmetric tensor $$B_{xy}={1\over 2}\cosh 2r\;\;\;\;\;
\;\; \Phi=0 \eqn\apq$$
The sigma model action that is described by $SL(2,R)/U(1)$
WZW coset model with level $k$
\Ref \wit{E. Witten \journal Phys.Rev. & D44 (91) 314
} (to one loop order)
is $${k\over 2\pi} \int d^2 \sigma (\partial_+r\partial_-r
\pm \tanh ^2 r \partial_+ t\partial_-t)  \eqn\witac$$with the dilaton
field $$\Phi=-\ln\cosh^2 r\eqn\wrt$$
 Alternatively it can be derived with
$\coth$ instead of $\tanh$, then    the dilaton is
                $\Phi=-\ln\sinh^2 r$ \Ref \giveo {A. Giveon,
 preprint LBL-30671}.
\Ref\kir{E.B. Kiritsis, \journal Mod.Phys.Lett &A6 (91) 2871}
(This is obtained by gauging the vector $U(1)$ rather than the
axial $U(1)$ gauge.) The
 $\pm$ sign accounts for the possibility to gauge either a compact
or non-compact  $U(1)$ subgroup.
The level is actually a multiplicative constant of the
metric and the anti-symmetric tensor,
and  can be absorbed
in the definitions of the coordinate, but in our next two cases we shall
leave it as an  overall factor.

Now,  consider the  background that contains
the coset model $SL(2,R)/U(1)$ and a decoupled
free compactified scalar
 field which we denote by $Z$ (throughout this paper $Z$ will
              denote a compactified scalar field):
$$d^2S=kd^2 r-k\tanh^2r d^2t+d^2Z,\;\;\;\;\;\;\;\;\;\;
 \Phi=-\ln\cosh^2 r \;\;\;\;\;\;  B_{\mu\nu}=0
\eqn\aelev$$
Now we can apply the $O(2,2)$ symmetry
transformations
 on $G_{tt}, G_{zz}, \Phi$.
 First
 we multiply $G_{zz}$
by $\beta^2=k\coth^2\alpha$ ($\ie$  rescale $Z\to \beta Z$) and
then apply the transformation
 with $\Omega$ given in
 $\ett$.
 For $R,S$
we choose the following $O(1,1)$ matrices:
$$S=\left (\matrix {\cosh\alpha &\sinh\alpha\cr
                    -\sinh\alpha&-\cosh\alpha\cr}\right)\eqn\bff$$
$$R=\left (\matrix {\cosh\alpha &\sinh\alpha\cr
                    \sinh\alpha&\cosh\alpha\cr}\right)\eqn\bfe$$
with $\alpha$ an arbitrary constant.
We obtain the following solution:
$$\tilde G_{rr}=k,\;\;\;\; \tilde G_{tt}=-k\cosh^{-2}
\alpha\sinh ^2 r
,\;\;\; \tilde G_{zz}=k^{-1}\sinh^{-2}\alpha\cosh^2 r \eqn\caa$$
$$\tilde B_{tz}={1\over 2}\cosh^{-1}
\alpha\sinh^{-1}\alpha\cosh 2r+{\cosh 2\alpha\over
2\sinh\alpha\cosh\alpha}
\;\;\;\;\;\; \tilde\Phi=0\eqn\cab$$   Now we rescale
  $t\to\cosh\alpha t$ and $z \to k\sinh^\alpha Z$
and subtract a constant anti-symmetric $2\times 2$ matrix
from the anti-symmetric tensor.
We see that the twisted solution is exactly the sigma-model
$\aqq$ that is described by the $SL(2,R)$ WZW model.

We want to note that the background
$$d^2S=kd^2r-k\cot^2 d^2 t+d^2 Z,
 \;\;\;\;\;\Phi=-\ln\sinh^2r\eqn\lm$$
can be obtained also by applying symmetry transformation on the
background   \aelev. The  matrices
$$S=R=\left (\matrix {-1&0\cr0&1\cr} \right )\eqn\dfsb$$
transform $G_{tt},G_{zz}$ to $G_{tt}^{-1},G_{zz}^{-1}$, respectively.
(So after the transformation we rescale $t\to kt$. Notice that
  we cannot transform from coset with level $k$ to
level $k^{-1}$).

We can obtain the $SL(2,R)$ WZW action $\aqq$ also by twisting
the following background
$$d^2 S=kd^2r+k\coth^2 rd^2 t-d^2 Z,\;\;\;\;\;\;\;\Phi=-\ln\sinh^2
 r \eqn\yyu$$ with the matrices $R,S$ given in \bfe\bff.

By reversing the $O(2,2)$ transformations we have used, we can
twist  the $SL(2,R)$ WZW action and
 obtain all the $SL(2,R)/U(1)$ coset models with additional
 free scalar  field.

In our second example we repeat this procedure
and obtain  $SU(2)/U(1)$
 WZW coset model from $SU(2)$ WZW action.   When the group elements
of $SU(2)$ are parameterized by $e^{i{1\over 2}\phi_L\sigma_3}
e^{i\theta \sigma_1}e^{i{1\over 2}\phi_R\sigma_3}$
the $SU(2)$ WZW action with (integer) level $k$
is
$$S_{SU(2)}={k\over 8\pi}\int d^2\sigma (4\partial_+\theta\partial_-
\theta
+\partial_+\phi_L\partial_-\phi_L +\partial_+\phi_R\partial_-\phi_R
+2\cos 2 \theta\partial_-\phi_L\partial_+\phi_R)\eqn\d$$
After defining the  fields  $x,y$ as in \aqw,
 the action becomes
 $$S_{SU(2)}=
{k\over 2\pi}\int d^2\sigma (\partial_+\theta\partial_-\theta
+\cos^2 \theta \partial_+x\partial_-x
+\sin ^2\theta \partial_+y\partial_-y    $$$$
+{1\over 2}\cos 2\theta(\partial_+x\partial_-y- \partial_-x\partial_+y))
\eqn\dd$$
  We start with the solution that include  the
coset $SU(2)/U(1)$ \Ref\rab{K. Bardakci, M. Crecimanno and E.
Rabinovici, "Parafermions From Coset Models", ILBL preprint (1990)}
 and the free field $Z$:
$$d^2S=k d^2\theta +k\tan^2\theta d^2\phi+d^2 Z, \;\;\;\Phi
=-\ln (\cos^2\theta)\eqn\suu$$
The $SU(2)/U(1)$ coset model can be obtained also
by analytic continuation
of the $SL(2,R)$ WZW action- by setting $r\to i\theta$ and changing
the sign of the level $k$. We can obtain the $SU(2)/U(1)$ action
with $\tan^2\theta$ in $\suu$ replaced by  $-\tan^2\theta$, or
 by $\pm\cot^2\theta$ with the dilaton $\Phi=-\ln\sin^2\theta$.
After we rescale  in $\suu$
 $Z\to \sqrt{k}\cot\alpha Z$,
  we apply the following $O(2)\times O(2)$ symmetry
$$S=\left (\matrix {\cos\alpha &\sin\alpha\cr
                    -\sin\alpha&\cos\alpha\cr}\right)\eqn\ft$$
$$R=\left (\matrix {\cos\alpha &\sin\alpha\cr
                    \sin\alpha&-\cos\alpha\cr}\right)\eqn\rgte$$
 In the new solution  we rescale
$\phi$ by $\cos^{-1}\alpha$ and $z$ by $k^{-1}\sin^{-1}\alpha$ and
add a constant anti-symmetric matrix to $B_{\mu\nu}$, as we did
in the previous example.
 The twisted solution is
then  the $SU(2)$ WZW model \dd.
The coset model with $\cot^2 \theta$ instead of $\tan^2\theta$
in $\suu$ can be  obtained by twisting the background in $\suu$
with $R=S=I$ where $I$ is the
$2\times 2$ unit matrix.

In the next example, we show that the coset model
${SL(2,R)\times U(1)\over U(1)}$ can also be obtained as a twisted
solution. This example was mentioned
by Sen [\senn]
in the context of p-brane solutions.
         Here  we bring this example for  completeness.
The coset model ${SL(2,R)\times U(1)\over U(1)}$ was derived in
\Ref\ish{N. Ishibashi, M. Li and A.R. Steif \journal Phys.Rev.Lett.
& 67 (92) 3336}
\Ref\garry{J.H. Horne, G.T. Horowitz and A.R. Steif \journal
Phys.Rev.Lett. & 68 (92) 568}.
 The background contains
$$d^2 S=kd^2 r-k{\sinh^2 r\over \cosh^2 r +\lambda}d^2 t +
k{\cosh^2 r\over \cosh^2 r+\lambda}d^2 x \eqn\gar$$
$$B_{tx}=k\sqrt{\lambda\over \lambda+1}
{\sinh^2 r\over \cosh^2 r +\lambda}
\;\;\;\;\;\; \Phi=-\ln (\cosh^2 r
+\lambda)\eqn\ggg$$
This background can be obtained by twisting the solution in \aelev.
Instead of rescaling $Z$ by $\sqrt {k}\coth\alpha$ as we did to obtain
\caa\cab,
we first rescale $Z$ by an arbitrary constant $q$, so $Z\to \sqrt{k}
q Z$.
 Now we apply the
$O(1,1)\times O(1,1)$ transformations with the  matrices $R,S$
 in \bff\bfe. The twisted background is the same as \gar\ggg
with $$\lambda={\tanh^2\alpha\over q^2-\tanh^2\alpha}\eqn\bvb$$
 (up to trivial rescaling of $t,Z$).

The Euclidean version of the background in \gar,\ggg is obtained by
twisting the background
$$d^2S= kd^2r+k\tanh^2r +d^2Z,\;\;\;\;\;\;\;\;
\Phi=-\ln\cosh^2 r\eqn\vv$$
 Similarly, the coset ${SU(2)\times U(1)\over U(1)}$ can be obtained
as a twisted solution. The procedure  is similar and therefore  we
do not bring it here.

 We see that all the coset models with $SL(2,R),SU(2)$ we discussed,
 can be obtained directly by $O(2,2)$ symmetry transformations
from the respective (ungauged) WZW models.

Now let us consider more complicated coset models.
It was shown \Ref\hor{P. Horava \journal Phys.Lett & B278 (92)
101}
 that  the coset model
$SL(2,R)_{k_1}\times SL(2,R)_{k_2}
/U(1)^2$ where  $k_1,k_2$  are the levels of the two groups,
can be described by the following
 background (here we absorb the levels in the coordinates):
$$d^2S= d^2 r_1+d^2 r_2 +{\cosh^2  ({r_1\over \sqrt{k_1}})
\sinh^2 ({r_2\over \sqrt{k_2}}) \over \Delta}
d^2\theta-{\sinh^2({r_1\over\sqrt{ k_1}})
\cosh^2({r_2\over \sqrt{k_2}})\over \Delta} d^2 t\eqn\dtt$$
$$B_{t\theta}={\sinh^2({r_1\over \sqrt{k_1}})
\sinh^2({r_2\over \sqrt{k_2}})\over \Delta},\;\;\;
\;\; \Phi=-\ln (\Delta)\eqn\dttt $$
with  $$\Delta= \cosh^2({r_1\over \sqrt{k_1}})
\cosh^2({r_2\over \sqrt{k_2}})
-\gamma^2\sinh^2({r_1\over \sqrt{k_1}})
\sinh^2({r_2\over \sqrt{k_2}})\eqn\dtf$$ and $\gamma$ is  a constant.
   Let us show
that  this background
  can be obtained by twisting the background that contains two decoupled
$SL(2,R)/U(1)$ coset models with levels $k_1,k_2$.
We start with the following solution:
$$d^2 S= d^2 r_1-\tanh^2({r_1\over \sqrt{k_1}}) d^2t
+d^2 r_2  +\cot^2({ r_2 \over \sqrt{k_2}})   d^2\phi\eqn
\wsl$$ $$B_{\mu\nu}=0,
\;\;\;\;\;\;\Phi=-\ln \cosh^2({r_1\over \sqrt {k_1}}) -
\ln\sinh^2({r_2\over \sqrt {k_2}})\eqn\slt
$$
First rescale $\phi\to q\phi$.
 Now we apply the $O(1,1)\times O(1,1)$
 symmetry transformation on $G_{tt},G_{\phi\phi},\Phi $
 with the matrices $S,R$ given in \bff,\bfe.
 we obtain the solution in \dtt,\dttt, with $\gamma$ replaced by
$q^2\tanh^2\alpha$.

 In the previous examples we have shown that coset models of a group $G$
gauged by $U(1)$ subgroup
 could be obtained by symmetry transformations from the (ungauged) WZW
model of the same group $G$.
 For example, both the cosets ${SL(2,R)\over U(1)}\times U(1)$
and ${SL(2,R)\times U(1)\over U(1)}$ and ${SL(2,R) \times SL(2,R)
\over U(1)^2}$ could be obtained directly
by twisting
$SL(2,R)$ WZW model, or  $SL(2,R)\times SL(2,R)$ in the latter.

 In the next example we demonstrate a different case.

Let us consider the $SL(2,R)\times SU(2)/U(1)^2$ coset model.
One way to write this coset is by analytic continuation of $r_1$
 to $i\theta$ in the background in \dtt\dttt,
followed by a change in sign of $k_1$.
Here, clearly we can obtain this model by twisting the
background that contains $SL(2,R)/U(1)$ and $SU(2)/U(1)$.
A different gauging (or gauge fixing) leads to quite a different
background, as was shown
 by Nappi and Witten \Ref\nappi{C.R. Nappi and E. Witten,
"A closed Expanding Universe in String Theory", preprint IASSNS-
HEP-92-38}.
 This describes the following background:
 $$d^2S= k_1d^2\theta_1-k_2d^2\theta_2 +{2\cos^2\theta_1 \cos^2\theta_2
(1+\sin\gamma)\over \Delta} d^2\phi_1 $$$$
+{2\sin^2\theta_1\sin^2\theta_2 (1-\sin\gamma)\over \Delta}d^2\phi_2
 \eqn\wee$$
$$B_{\phi_1\phi_2}={\cos 2\theta_2 -\cos2\theta_1 +\sin\gamma
(1-\cos 2\theta_1\cos 2\theta_2)\over \Delta}\;\;\;\;\;
\Phi =\ln\Delta\eqn\weet$$
with $$\Delta=1-\cos 2\theta_1\cos 2\theta_2+\sin\gamma (\cos 2
\theta_2-\cos 2\theta_1)\eqn\wett$$  where $k_1,k_2$ are the
levels of the $SL(2,R), SU(2)$ groups respectively, and we
absorbed them in $\phi_1,\phi_2$.
This  background can be obtained by twisting a background
that contains two $SU(2)/U(1)$ WZW coset  models one with level
$k_1$ and the other with $-k_2$. For unitarity we must restrict
 ourselves to integer $k_1$, although in
  $\wee$ $k_1$ can be    none integer.

We take the following solution:
$$d^2S= k_1d^2\theta_1 +\tan^2\theta_1d^2\phi_1 -k_2d^2\theta_2
+\tan^2\theta_2 d^2\phi_2\eqn\srt$$
$$B_{\mu\nu}=0\;\;\;\;\;\;\;\;
\Phi=-\ln\cos^2\theta_1-\ln\cos^2\theta_2
\eqn\stry$$
Applying the $O(2)\times O(2)$ symmetry transformations
on $G_{\phi_1\phi_1},G_{\phi_2
\phi_2}$ with the matrices $R,S$ given in \ft,\rgte, then
rescaling $\phi_2,\phi_2$ and adding a constant anti-symmetric
matrix to $b_{\mu\nu}$,  we obtain
the solution \wee\weet, with $\gamma$ replaced by $\cos 2\alpha$.
Hence evidently,
the symmetry transformed the solution to  a different conformal
theory.

 We can conclude the following:
Denote by $G_1$ the group $SL(2,R)$ and by $G_2$ the group
$SU(2)$. Then
$G_i\times G_j /U(1)^2$ WZW coset models can be obtained by twisting
the backgrounds  of $G_k/U(1)\times G_l/U(1)$ or directly from
$G_k\times G_l$ WZW, where  $(i,j)$ are not necessarily $(k,l)$.

Finally, we want to comment about the generality of our results.
When gauging a $U(1)$ subgroup in  WZW action, one can gauge
either a vector $U(1)$ or axial $U(1)$. Therefore, the gauged
action has a residual global $U(1)$ symmetry. This means that
one could apply the $O(d,d)$ transformation.
 However, when gauging a non-abelian
subgroup, it is possible to gauge only    vector gauge
 and thus the gauged
action will not necessarily be independent of some coordinates.
So we might conjecture that coset models involving only $U(1)$
gauging can be obtained by symmetry transformations from
(ungauged) WZW actions.

\chapter{ Twisted Solutions of
Charged Black Holes in Four Dimensions}

In this section we obtain  backgrounds that describe
 charged black holes in four dimensions by twisting a conformal
background. As we mentioned, we are guaranteed that the twisted
solutions correspond to  exact conformal theories.

Closed string theories with  gauge fields in their massless spectrum
can be constructed by introducing free bosons (or equivalently
free fermions) on the world sheet\Ref \gros {D.J. Gross, J.A. Harvey,
E. Martinec and R. Rohm \journal Phys.Rev.Lett. & 54, (85) 502}
\Ref\cal {C.G. Callan, D. Friedan, E.J. Martinec and M.J. Perry
\journal Nucl.Phys. & B262 (85) 593}. Bosonic string models with
gauge fields can be described by the set of the bosonic fields
 $X^{\mu},X^I$, where the $X^{\mu}$
 describe the  target space coordinates,
and the $X^I$, which are compactified fields, realize the Kac-Moody
currents of the gauge group. Unlike in the heterotic strings, where
 only chiral
bosons are included,  in our model we introduce bosons with both
left and right chirality. In such models there is a separate gauge
symmetry for the left and the right currents. The model is
described by the following
 sigma-model action:
$$S={1\over 2\pi}\int d^2\sigma ((G_{\mu\nu}+B_{\mu\nu})\partial_+
X^{\mu}\partial_-X^{\nu} +A^I_{\mu}\partial_+X^{\mu}\partial_-Z_I
+\tilde A^I_{\mu}\partial_-X^{\mu}\partial_+Z_I $$$$
+\partial_+Z_I
\partial_-Z_I)-{1\over 8\pi}\int d^2\sigma hR^{(2)}\Phi\eqn\ch$$
where the fields $Z^I$ are compactified bosonic fields, the index  $I$
corresponds to the generators of the gauge group (and the last term
is the usual dilaton part in the action).
In the example we show, $A_{\mu}$ is an abelian gauge field, and the
action contains only one  $Z$ field.

The procedure we use in the following: We twist a conformal model
and then identify the new background with \ch . (We note that
one could also apply here the symmetry
transformations that were shown
for the heterotic strings in [\hassan].)
we start with the background that contains the cosets
$SL(2,R)/U(1)$ with the level $k_1$,
 $SU(2)/U(1)$ with the level $k_2$
  and a free compactified field $Z$.
This is described by the following background:
$$d^2S=k_1 d^2 \hat r-k_1\coth^2 \hat r +k_2d^2\theta
+k_2\cot^2\theta
d^2\phi+d^2Z      \eqn\err$$
$$B_{\mu\nu}=0\;\;\;\;\;\;\Phi=-\ln\sinh^2r-\ln \sin^2\theta\eqn\ert$$
As we explained in the previous section, this coset can be obtained by
twisting $SL(2,R)\times SU(2)$ WZW action.
We absorb the levels by redefining
$\hat r\to \sqrt{k_1\over k_2}\hat r$
$t\to \sqrt{k_1\over k_2}t$ to obtain an overall factor of $k_2$.
 First we rescale $Z$ by a constant, $Z\to q\sqrt{k_2}Z$.
Now we apply  $O(3,3)$
symmetry transformations on $G_{tt},G_{\phi\phi},
G_{zz}$.
In order to simplify, we do it in two stages.
First  we twist only $G_{\phi\phi}$ and $G_{zz}$,
 then we twist the
obtained solution once again.
In the first sage
  we take the following $O(1,2)$ matrices for $S$ and $R$:
$$S=\left(\matrix{-1&0&0&\cr 0&\cos\alpha&-\sin\alpha\cr
  0&\sin\alpha&\cos\alpha\cr}\right)\eqn\qq$$
$$R=\left(\matrix{-1&0&0&\cr 0&\cos\alpha&\sin\alpha\cr
  0&\sin\alpha&-\cos\alpha\cr}\right)\eqn\qq$$
This transforms the background \err,\ert
to the following  background:
$$ \tilde G _{tt}=G_{tt},\;\;\;  \tilde G_{\phi\phi}=
k_2\cos^{-2}\alpha(\cot^2\theta+q^2\tan^2\alpha )^{-1}
,$$$$
\tilde G_{zz}=k_2^{-1}\sin^{-2}\alpha
 (\tan^2\theta+q^{-2}\cot^2\alpha)^{-1}
\eqn\lk$$
$$\tilde B_{z\phi}=\tan\alpha {\cos^2\theta-q^2\sin^2\theta\over
1+(q^2\tan^2\alpha -1)\sin^2\theta}  +c
\eqn\klt$$
the metric is diagonal and other components
 of  the anti-symmetric tensor vanish.
 We also added a constant anti-symmetric matrix to $B_{\mu\nu}$
with only two components $B_{z\phi}=c, B_{\phi z}=-c$.
As can be seen from \dilat, the dilaton field can be calculated
after the second stage.

  After rescaling $Z\to k_2 Z$,
let us  denote $\tilde G_{tt},
  \tilde G_{\phi\phi},\tilde G_{zz}$
 by $g_1,g_2,g_3$, respectively.
Now we apply another  symmetry transformation, with
$$S=\left (\matrix{-\cosh\beta&-\sinh\beta&0 \cr
                   \sinh\beta&\cosh\beta&0 \cr
                    0&0&1\cr}\right )\eqn\klt$$
$$R=\left (\matrix{\cosh\beta&\sinh\beta&0 \cr
                  \sinh\beta&\cosh\beta&0 \cr
                    0&0&1\cr}\right )\eqn\klf$$
 We obtain the following metric,  written in matrix form:
 $$G=\Delta^{-1}\left (\matrix{1
  &0& \sinh\beta \tilde B_{\phi z} \cr
0&g_1g_2  & 0\cr
 \sinh\beta \tilde B_{\phi z}& 0&
 g_3\Delta+\sinh^2\beta \tilde B_{\phi z}^2\cr}\right )
  \eqn\jj$$
  the anti-symmetric tensor
$$B=\Delta^{-1}
\left (\matrix{0&\sinh\beta\cosh\beta(g_1+g_2)&
0\cr - \sinh\beta\cosh\beta(g_1+g_2)&0&\cosh\beta
\tilde B_{\phi z}g_1 \cr
 0&-\cosh\beta \tilde B_{\phi z}g_1&0\cr}\right)\eqn\anti$$
and the dilaton field
$$\Phi=-\ln(\sinh (\hat r\sqrt{k_2\over k_1})
\cosh^2 (\hat r\sqrt{k_2\over k_1}) \sin\theta\cos^2\theta)
+{1\over 2}\ln (g_1g_2g_3 )-\ln\Delta
\eqn\dil$$
   where
 $$\Delta=(\cosh^2\beta  g_1+\sinh^2\beta g_2)\eqn\ds$$
Finally, we  define $r=
\cosh^2(\sqrt{k_2\over k_1}
\hat r)$ and rescale $t,\phi$ by $\cosh^{-1}\beta,\sinh^{-1}\beta$
respectively.
 Identifying the twisted solution with the action  \ch, we obtain
 the following background:

$$d^2S=  -{(r-1)(1+Q\sin^2\theta)\over \Sigma}d^2 t+
{k_1d^2 r\over k_2(r-1)r} + {r\sin^2\theta\over \Sigma}d^2\phi +
d^2\theta \eqn\fin$$
$$B_{t\phi}=\sinh^2\beta
{r+\cos^2\alpha\sin^2\theta-r\sin^2\theta(\cos^2\alpha
-Q)\over \Sigma} \eqn\cqmn$$
$$A_t=e{
1+c\cot\alpha +(cQ\cot\alpha-1-q)\sin^2\theta \over\Sigma}
(r-1)\eqn\elds$$
$$A_{\phi}=                    e
{1+c\cot\alpha +(cQ\cot\alpha-1-q)\sin^2\theta \over\Sigma}r
\eqn\eldp$$      and the dilaton field
$$\Phi=-\ln\Sigma \eqn\ddil$$
where $Q=q^2\tan^2\alpha -1$,  $e= \tanh\beta\tan\alpha$
$$\Sigma=r+(Q-b)
r\sin^2\theta +b\sin^2
\theta\eqn\sigma$$ and   $b=\tanh^2\beta\cos^{-2}\alpha$.
 $G_{zz}$ gives rise to an additional scalar field with the
vertex operator $$V=\partial Z \bar \partial Z e^{-k_t t+
k_r r+ k_{\theta}\theta +k_{\phi}\phi}\eqn\vetr$$

The electric and the magnetic fields are obtained
by calculating the electromagnetic tensor $F_{\mu\nu}=
\nabla_{\mu} A_{\nu}- \nabla_{\nu}A_{\mu}$.

Now let us discuss the metric \fin.
First of all, one can see
that there is an event horizon at $r=1$, as long as there is
no singularity at $r=1$. We shall consider this soon.
{}From the definition of $Q$ we see that  $Q> -1$
when both  $q,\tan \alpha\ne 0$
($q=0$  means that the original solution
does not contain the free field $Z$).  So let us restrict ourselves
from now on
to the solutions with $\tan\alpha\ne 0$.   $G_{tt}$ then vanishes
only at $r=1$.

 The metric has a singularity only at $\Sigma=0$ (for $Q>-1$).
This can be seen by calculating the curvature tensors and
the scalar curvature. (The expressions are very long and we
do not bring them here.)   For $Q-b>-1$ the singularity
occurs when $r=\sin^2\theta =0$. Therefore    the singularity
is surrounded by the event horizon. In the other case, when
$Q-b<-1$ the singularity is at $r={b\sin^2\theta\over
(b-Q)\sin^2\theta-1}$ and thus
  $r=1$ would not be an event horizon.
 Hence in order to describe
 black holes, we must restrict $b,Q$ so that $Q-b>-1$.

The Einstein metric $G^E_{\mu\nu}$
is obtained by rescaling the metric $G_{\mu\nu}$
of  the sigma-model
with the dilaton field. In four dimensions
$G^E_{\mu\nu}=e^{-\phi}G_{\mu\nu}$.
 In our case the Einstein metric
  is singular at $\Sigma=0$
only and thus the restriction $Q-b>-1$ is still valid.

In order that our metric describes a black hole we should
verify that the area of the event horizon is finite.
The area of the event horizon is given by
$$A_{horizon}=\int d\theta d\phi\sqrt{G_{\theta\theta}
G_{\phi\phi}}\eqn\hori$$ at $r=1$,
where $\theta$ is from 0 to $\pi$
and $\phi$  from 0 to $2\pi$. We can see that
 the condition for the area of the event horizon to be finite
 is $Q>-1$.
Thus, for $Q-b>-1$
 the solution $\fin$ describes a class of axi-symmetric
black-holes
 with electromagnetic and axionic fields.
In particular, one can choose $Q=b$, and then the
singularity is at $r+Q\sin^2\theta=0$. This is the same type
of singularity that occurs in the Kerr Solution that leads to
a ring-type singularity.   The solution with $Q=b$
was already obtained by us from the WZW coset model
$SL(2,R)\times SU(2)\times U(1)/U(1)^2$\Ref\gershon{D. Gershon,
"Exact
Stringy Charged and Uncharged Black Hole Solutions in Four
Dimension",  Proceeding of The 1992 ICTP Summer School in High
Energy Physics and Cosmology, World Scientific publishing Co.}
 (up to additional anti-symmetric constant matrices).

The asymptotic shape of these solutions, obtained as
$r\to\infty$ is
$$d^2S=-{1+Q\sin^2\theta\over 1+(Q-b)\sin^2\theta}
d^2t+{1\over r^2}(d^2r+{\sin^2\theta d^2\phi\over 1+(Q-b
)\sin^2\theta}+d^2\theta)\eqn\asym$$
For $Q=b$ the space turns  to be spherically symmetric
at infinity.

Finally, from $\elds,\eldp$ we can  see that there are both
 electric and magnetic fields.
The electric and the magnetic charges can be calculated, as
conserved charges of the effective actions. The electric
charge is obtained by $$q_{E}=\int e^{-\phi+\varphi}{}^* F
d^2 S \eqn\elel$$ where $\varphi=\ln\sqrt{G_{zz}}$ and
the integral is over a 2-sphere at
infinity ($^* F$ is F-dual). In our case we obtain:
              following expression:
$$q_E=4\pi e\int_0^{\pi} d\theta {1+c\cot\alpha+(cQ\cot\alpha
-1-q)\sin^2\theta\over \sqrt{1+(Q-b)\sin^2\theta}}\eqn\wqnmio$$
The magnetic charge is $$q_{mag}=\int F d^2 S =0\eqn\malh$$
namely, the black hole carries only electric charge.

However,  the effective action is invariant under the
transformation $F\to ^*F$
 \Ref\stro{D. Garfinkle, G.T. Horowitz and
A. Strominger \journal Phys.Rev. & D 43 (91) 3140}.
 Thus one can obtain also black hole
solutions with magnetic charge.

\chapter{Summary and Remarks}

In this paper we have shown that well known coset models
 of gauged $U(1)$ or $U(1)^2$ subgroups
can be obtained by  $O(d,d)$ symmetry transformations from
 (ungauged) WZW action.
  In many cases, both the background that
was twisted and the twisted solution described equivalent
conformal theories and thus the $O(d,d)$ symmetry was
a duality transformation. In another case we demonstrated that
the symmetry  transformed from the coset $SL(2,R)
\times SU(2)/U(1)^2$ to the coset
${SU(2)\over U(1)}\times {SU(2)\over
U(1)}$.
 This leads to a conjecture that gauging WZW models by
 $U(1)^n$ subgroup, can be done by
$O(d,d)$ transformations from
(ungauged) WZW actions.


Finally, we found a class of solutions that describe electrically
charged
black holes in four dimension, and that correspond
to exact conformal field
theories.
A class  of magnetically
charged black holes can be obtained by the symmetry transformation
from the electromagnetic tensor to its dual.

\vskip 2 true cm

\ack
I want to thank A. Sen for encouraging me in the summer
to study the
$O(d,d)$ symmetries.
\vskip 2 true cm

\centerline {NOTE ADDED}

After the completion of this work, we were notified that some
of the results appearing in out paper were also obtained in a
recent paper
by A.F. Hassan and A. Sen\Ref\sh {A.F. Hassan and A. Sen,
"Marginal Deformations of WZNW and Coset Models From O(d,d)
Transformations", preprint TIFR-TH-92-61 (hepth/9210121)}.
 In particular, the possibility to
obtain some coset models  by  $O(d,d)$
transformation from  (ungauged) WZW models.
\refout
\end\bye